\documentstyle[preprint,aps]{revtex} 
\tightenlines
\begin{document} 
\draft

\title{\begin{flushright}
          {\small IFT-P. 043/98 \,\, gr-qc/9806093}
       \end{flushright}
     Interaction of Hawking radiation with static 
sources outside a Schwarzschild black hole} 


\author{Atsushi Higuchi}
\address{Institut f\"ur theoretische Physik, Universit\"at Bern,\\
          Sidlerstrasse 5, CH--3012 Bern, Switzerland\\ and\\
         Department of Mathematics, University of York,\\ 
         Heslington, York YO1 5DD, United Kingdom}
\author{George E.A. Matsas}
\address{Instituto de F\'\i sica Te\'orica, 
         Universidade Estadual Paulista,\\
         Rua Pamplona 145,
         01405--900--S\~ao Paulo, S\~ao Paulo,
         Brazil}
\author{Daniel Sudarsky}
\address{Instituto de Ciencias Nucleares,
         Universidad Nacional Aut\'onoma de M\'exico,\\
        A. Postal 70/543, M\'exico D.F. 04510,
       M\'exico}
\date{\today}
\maketitle 


\begin{abstract}
We show that the response rate of (i) a static source interacting with 
Hawking radiation of massless scalar field in Schwarzschild
spacetime (with the Unruh vacuum) and that of (ii) a 
uniformly accelerated source with the same proper acceleration
in Minkowski spacetime (with the Minkowski vacuum) are equal. 
We show that this equality will not hold if the Unruh vacuum is replaced
by the Hartle--Hawking vacuum.  It is verified that the source responds to
the Hawking radiation near the horizon as if it were at rest in a thermal
bath in Minkowski spacetime with the same temperature.  It is also verified
that the response rate in the Hartle--Hawking vacuum approaches that in
Minkowski spacetime with the same temperature far away from the black hole. 
Finally, we compare our results with others in the literature.
\end{abstract} 

\narrowtext 

\section{Introduction}
\label{sec:Introduction}

Recently we analyzed the emission and absorption of
``zero--energy particles" by a static source interacting with
Hawking radiation outside a Schwarzschild black hole
\cite{HMS2}.  It was found that the total response rate of a
point--like static scalar source in the Unruh vacuum is given by
\begin{equation} 
R_{\rm tot} = \frac{q^2 a}{4\pi^2}\,,
\label{main} 
\end{equation} 
where $q$ is the coupling
constant between the source and the massless scalar 
field, and $a$ is the proper acceleration of the source.  The
remarkable fact about this result is that Eq.\ (\ref{main})
also corresponds to the total response rate of a uniformly
accelerated source for the massless scalar field in  Minkowski spacetime 
provided that the initial quantum state 
is the Minkowski vacuum.  (In fact, according to inertial
observers, Eq.\ (\ref{main}) is associated with the emission
rate of finite--energy Minkowski particles while according
to coaccelerated observers it is associated with the
emission and absorption of zero--energy Rindler particles
\cite{HMS}.)  Thus, an equality between the
behavior of static sources in Schwarzschild spacetime
(with the Unruh vacuum) and uniformly accelerated sources in
Minkowski spacetime (with the Minkowski vacuum), 
concerning their emission and absorption rates, was
found.  Here we analyze in detail  some related points that were
not discussed in Ref.\ \cite{HMS2}, provide some 
consistency checks of our results and demonstrate their compatibility
with related results in the literature.  
The paper is organized as follows.  In
Sec.\ \ref{sec:General}, we  review the general formalism for
computing the response rate of classical sources for a massless scalar
field in static
spacetime.  In Sec.\ \ref{sec:Rindler}, we analyze the case of a
static point--like source in the Rindler wedge -- i.e.  a
uniformly accelerated point--like source in Minkowski
spacetime -- and show that its total response rate (with
the initial quantum state 
being the Minkowski vacuum) is given by Eq.\ (\ref{main}).  Next, in
Sec.\ \ref{sec:Minkowski}, we consider a static
source immersed in a thermal bath in Minkowski spacetime
and calculate its response rate for later use.  In Sec.\ \ref{sec:bhtoy},
we consider a toy model for a static source outside a
static black hole characterized by a simplified
gravitational effective potential and calculate its response rate assuming
that the initial quantum state 
is the Unruh vacuum.    In Sec.\ \ref{sec:bhS}, the
response rate is calculated with 
the true  Schwarzschild effective potential.
The results found are compared with those
obtained in Sec.\ \ref{sec:Rindler}.  In particular, it is shown that the
total response rate here is also given by
Eq.\ (\ref{main}).  This equivalence is our main result
\cite{HMS2}.  In Sec.\ \ref{sec:bhS2}, we calculate the
response rate replacing the Unruh vacuum by the
Hartle--Hawking one and show that Eq.\ (\ref{main}) does not hold. 
In Sec.\ \ref{sec:literature}, we 
discuss  the case where the
source approaches the horizon and the case where it is
far away from the black hole using the method described 
in Refs.\ \cite{C,CSD} and show that the results 
agree with the suitable limit of the one obtained in 
Sec.\ \ref{sec:bhS}, i.e., Eq.\ (\ref{main}).  In Sec.\ \ref{sec:discussions},
we discuss our results.
We will use natural units $\hbar=c=G=k_B=1$ throughout
this paper.


\section{General Formalism}
\label{sec:General}

It is well known that field theory 
quantized in globally--hyperbolic 
spacetime possessing a global timelike Killing field 
admits a unique vacuum state and the corresponding unique ``particle 
interpretation"
(under certain technical conditions) \cite{Ash,Ber}.
This is so because  the use of the time parameter corresponding to the Killing
field allows us to distinguish, in a natural way, 
between positive and negative
frequency modes.
This is the case in 
globally--hyperbolic static spacetime
described by the metric
\begin{equation}
ds^2 = f({\bf x})dt^2 - h_{ij}({\bf x})dx^i dx^j,
\label{ST}
\end{equation}
(under certain technical conditions).
We consider emission of particles in these spacetimes
by classical 
static scalar sources $J({\bf x})$ coupled to a massless scalar field. 
The response of a classical source in the vacuum is entirely due to
spontaneous emission.  If the source is static, then this vanishes (unless
there are severe infrared divergences).  However, if the static source is
in a thermal bath, the absorption and induced emission also contribute to
the response rate.  Now, the static source interacts only with zero--energy
modes and Planck's distribution formula diverges at zero energy.
It will turn out that this makes the rates of absorption and induced emission
nonzero.  Thus, the static source responds with a finite probability to 
thermal baths for the cases we consider in this paper.

In order to avoid the appearance of intermediate indefinite 
results due to the divergence mentioned above,
we will introduce oscillation as a regulator.
Thus, we consider at this point a source of the form
\begin{equation}
j_{\omega_0}(t,{\bf x}) = \sqrt 2 J({\bf x})\cos\omega_0 t 
\label{j}
\end{equation}
and take the limit $\omega_0 \to 0$ at the end. The factor  $\sqrt 2$
has been introduced to keep the time average 
$\langle |j_{\omega_0 }(t,{\bf x})|^2 \rangle_t$
equal to $|J({\bf x})|^2$. This makes 
$j_{\omega_0 }(t,{\bf x})$ equivalent to $J({\bf x})$ 
in the limit $\omega_0 \to 0$ because 
the response rate at the lowest order
is proportional to the square of (the Fourier transform of) the source.
We will be interested in the point source where
\begin{equation} 
 J({\bf x}) = q \delta ({\bf x} - {\bf x}_0)/ \sqrt{h}
\label{J}
\end{equation}
with $q$ being the coupling constant, ${\bf x}_0$ being 
the position of the source and 
$h({\bf x}) \equiv {\rm det}\,[h_{ij}({\bf x})]$. 
With this definition, we have
\begin{equation}
\int_{\Sigma_t} d\Sigma \, J({\bf x}) = q
\end{equation}
for any Cauchy surface $\Sigma_t$ with constant $t$.

Let us  consider the coupling of our classical 
source $j_{\omega_0}(t,{\bf x})$
to a massless real scalar field $\Phi$, which is described by the action,
$$
S = \int d^4 x \sqrt{fh}\,\left( {1 \over  2}
\nabla^{\mu}\Phi \nabla_{\mu} \Phi
 +j_{\omega_0} \Phi\right) .
$$ 
Let 
\begin{equation}
u_{\omega{\bf s}\lambda}(x)
= \sqrt{\frac{\omega}{\pi}}\,
U_{\omega {\bf s}\lambda}({\bf x})\exp(-i\omega t) 
\label{Udef}
\end{equation}
with  frequency $\omega > 0$, and their complex
conjugates
$u_{\omega {\bf s}\lambda}(x)^{*}$ be solutions to 
$\Box u = 0$, 
where ${\bf s} = (s_1,\cdots, s_n)$ and 
$\lambda=(\lambda_1, \cdots, \lambda_m)$ are sets of
continuous and discrete quantum numbers, respectively,
for the complete set of modes. We have assumed $\omega$ to be 
continuous because this is the case in the spacetimes we study, 
and adopted it as one of the mode 
labels. The factor 
$\sqrt{\omega/\pi}$ has been inserted for later
convenience.  We orthonormalize these solutions
with respect to the Klein--Gordon inner product:
\begin{eqnarray}
i\int_{\Sigma_t} d\Sigma \, n^{\mu}\left( 
u_{\omega {\bf s}\lambda}^{*}
\nabla_{\mu}u_{\omega'{\bf s}'\lambda'}
- \nabla_{\mu}u_{\omega {\bf s}\lambda}^{*}\cdot 
u_{\omega' {\bf s}'\lambda'} \right) 
&=&
\delta(\omega - \omega')\delta({\bf s}-{\bf s}')\delta_{\lambda\lambda'},
\label{KG} 
\\
i\int_{\Sigma_t} d\Sigma \, n^{\mu}\left( 
u_{\omega{\bf s}\lambda}\nabla_{\mu}u_{\omega'{\bf s}'\lambda'}
- \nabla_{\mu}u_{\omega{\bf s}\lambda}\cdot 
u_{\omega'{\bf s}'\lambda'}\right)
&=&  
0, 
\end{eqnarray}
where $n^{\mu}$ is the future--pointing unit normal to the 
volume element of a Cauchy surface $\Sigma_t$. 
The in--field $\Phi^{\rm in}$ satisfying 
the free field equation $\Box \Phi^{\rm in} = 0$ can now be expanded as 
$$
\Phi^{\rm in}(x) = \sum_{\lambda} \int d\omega d^n{\bf s}
\left[ u_{\omega {\bf s}\lambda}(x) a^{\rm in}_{\omega {\bf s}\lambda} 
+ H.c. 
\right],
$$
where 
$a^{\rm in}_{\omega {\bf s}\lambda}$ 
and 
$a^{{\rm in}^\dagger}_{\omega {\bf s}\lambda} $ 
are annihilation and creation  operators, respectively,
which satisfy the usual commutation relations
\begin{equation}
[ a^{\rm in}_{\omega {\bf s}\lambda}, 
a^{{\rm in}\dagger}_{\omega' {\bf s}' \lambda'}] 
= 
\delta(\omega - \omega')
\delta({\bf s}-{\bf s}')
\delta_{\lambda\lambda'}.
\label{CR}
\end{equation}

Let the initial quantum state be the in--vacuum state 
$|0\rangle$ defined by
$a^{\rm in}_{\omega{\bf s}\lambda}|0\rangle = 0$ for all $\omega$,
${\bf s}$ and $\lambda$. 
The rate of spontaneous emission per 
total {\em proper} time $T \sqrt{f({\bf x}_0)}$, 
where $T\equiv 2\pi \delta (0)$ is the total 
{\em coordinate} time \cite{IZ},
with fixed ${\bf s}$ 
and $\lambda$ to lowest order is
\begin{equation}
R_{\rm sp}(\omega_0;{\bf s},\lambda)d^n{\bf s} = 
T^{-1} f({\bf x}_0)^{-1/2} \int{d\omega 
\left| \left\langle \omega {\bf s} \lambda 
\left|  \int d^4 x \sqrt{fh}\, j_{\omega_0} \Phi \, 
\right| 0 \right\rangle 
\right|^2 d^n{\bf s}} . 
\end{equation} 
By performing the integration with respect to $\omega$, we obtain
\begin{equation}
R_{\rm sp}(\omega_0;{\bf s},\lambda)d^n{\bf s} =
\frac{\omega_0}{\sqrt{f({\bf x}_0)}}
|\tilde{J}(\omega_0; {\bf s},\lambda)|^2 d^n{\bf s}, 
\label{Sprate}
\end{equation}
where 
$$
\tilde{J}(\omega_0;{\bf s},\lambda) = \int d^3{\bf x}\,
\sqrt{f({\bf x})h({\bf x})}\,
J({\bf x})
U_{\omega_0 {\bf s}\lambda}({\bf x}). 
$$

If the source is immersed
in a thermal bath of inverse temperature
$\beta$, the rates of absorption and {\em induced} emission
are both given by 
$R_{\rm sp}(\omega_0;{\bf s},\lambda)/(\exp\beta\omega_0 - 1)$.
Adding the absorption rate and the 
spontaneous and induced emission rates,
we find that the {\em response} rate 
for modes with fixed ${\bf s}$ and
 $\lambda$ is given by
$$
R(\omega_0;{\bf s},\lambda)
= \frac{\omega_0}{\sqrt{f({\bf x}_0)}} 
|\tilde{J}(\omega_0;{\bf s},\lambda)|^2 \coth (\beta\omega_0/2)  .
$$
In the case of interest here, i.e., for static sources, we take 
the limit $\omega_0 \to 0$ as explained above, obtaining 
\begin{equation}
R(0;{\bf s},\lambda)
= \frac{2 \beta^{-1}}{\sqrt{f({\bf x}_0)}}
|\tilde{J}(0;{\bf s},\lambda)|^2. 
\label{Maineq}
\label{important1}
\end{equation}
This is the general expression for the response rate  
of a static point source in a thermal bath interacting 
with a massless scalar field.
The total rate is obtained from (\ref{Maineq}) by integrating with respect to 
${\bf s}$ and summing over $\lambda $. Note that although 
the particle content of a field theory depends in general 
on the Killing field with respect to which the vacuum is
defined, the total response rate does not.


\section{Static source in the Rindler wedge}
\label{sec:Rindler}

We first review the computation of 
the response rate of a static source in the 
Rindler wedge \cite{Rind} (see Refs.\ \cite{HMS,RW}).  
This source corresponds to 
a uniformly accelerated source
in Minkowski spacetime. The Rindler wedge 
is the portion of 
Minkowski spacetime limited by $z > |t|$, 
where $(t,x,y,z)$ are the usual 
Minkowski coordinates.   These are related 
to the Rindler coordinates 
$(\tau,x,y,\xi)$  by  
$$
t  = a^{-1}e^{a\xi}\sinh a\tau, \;\;\;\;
z  = a^{-1}e^{a\xi}\cosh a\tau.
$$
In these coordinates, the line element of the Rindler wedge 
is written as
$$
ds^2 = e^{2 a \xi} (d\tau^2 - d\xi^2) -dx^2 - dy^2 . 
$$
We consider a point--like 
 source fixed in space with time--dependent magnitude 
 [see Eqs.\ (\ref{j}) and (\ref{J})]
\begin{equation}
j_{\omega_0} = \sqrt{2}\,q\cos\omega_0 \tau\,
\delta(\xi)\delta(x)\delta(y) 
\label{Rsource}
\end{equation} 
and take the limit $\omega_0 \to 0$ at the end.  
Note that $j_{\omega_0} $ describes
a  source with constant proper acceleration $a$. 

We will describe the free massless scalar field theory using
Rindler coordinates. For this purpose 
we look for positive--frequency
solutions to $\Box u_{\omega k_x k_y}=0$ 
with respect to the Killing field $\partial/\partial \tau$:
\begin{equation}
u_{\omega k_x k_y}(\tau,x,y,\xi) =\sqrt{\frac{\omega}{\pi}}\,
\psi_{\omega k_{\perp}}(\xi) \times 
\frac{e^{ik_x x + ik_y y - i\omega \tau}}{2\pi} . 
\label{Uinpsi}
\end{equation}
Here the $\psi_{\omega k_{\perp}}$ must satisfy
\begin{equation}
\left[ -\frac{d^2\ }{d\xi^2} + k_{\perp}^{2}e^{2a\xi}\right]
\psi_{\omega k_{\perp}}(\xi) = \omega^2\psi_{\omega k_{\perp}}(\xi),
\label{MObessel}
\end{equation}
and $k_{\perp} \equiv \sqrt{k_x^2 + k_y^2}$\,. 
We assume the Minkowski 
vacuum which corresponds to a thermal state 
of Rindler particles 
\cite{D,U,F}. This is the 
Fulling--Davies--Unruh (FDU) thermal bath 
characterized by a temperature  $\beta^{-1} = a/2 \pi$. 
The term $k_\bot^2 e^{2 a \xi}$
in (\ref{MObessel}) acts as an effective potential 
that is unbounded for the modes with nonvanishing 
transverse--momentum.  The solutions 
$\psi_{\omega k_{\perp}}(\xi)$ 
that are finite for $\xi \to +\infty$ are
\begin{equation}
\psi_{\omega k_{\perp}}(\xi) = 
C_\omega K_{i\omega/a}[(k_{\perp}/a)e^{a\xi}],
\label{PFRW}
\end{equation}
where $C_\omega$ is a normalization constant. In order to
determine it, we substitute normal modes (\ref{Uinpsi}) in 
the Klein--Gordon inner product (\ref{KG}), obtaining
\begin{equation}
(\omega + \omega')  
\int_{-\infty}^{+\infty} d\xi 
\psi_{\omega k_\perp}^* (\xi )
\psi_{\omega' k_\perp}  (\xi )
 = \frac{\pi}{\omega} \delta (\omega - \omega') .
\label{KGRW}
\end{equation}
Next we use the wave equation (\ref{MObessel}) 
to turn the integral in Eq.\ (\ref{KGRW}) into a surface term:
\begin{equation} 
\frac{1}{\omega - \omega'}\left[ \psi_{\omega' k_\perp} 
\frac{d \psi_{\omega k_\perp}^*}{ d\xi} 
-
\psi_{\omega k_\perp}^* 
\frac{d \psi_{\omega' k_\perp} }{ d\xi} \right]_{\xi \to -\infty}
=
\frac{\pi}{\omega} \delta (\omega - \omega') .
\label{aux1}
\end{equation}
Using (\ref{PFRW}) in (\ref{aux1}) and noting that
\begin{equation}
\lim_{r \to +\infty} \frac{\sin (\omega \mp \omega')r}{\omega \mp \omega'} 
= \pi \delta (\omega \mp \omega'),  \label{Delta}
\end{equation}
we obtain the normalization 
constant (up to a phase)
\begin{equation}
C_\omega = \sqrt{\frac{\sinh (\pi \omega/a)}{\pi a \omega}} .
\label{C}
\end{equation}
{}From (\ref{PFRW}) and (\ref{C}) we obtain
the normalized zero--energy modes
\begin{equation}
\psi_{0 k_{\perp}}(\xi) = a^{-1}
K_{0}[(k_{\perp}/a)e^{a\xi}]. \label{Zero}
\end{equation}

In fact, one can directly normalize $\psi_{0 k_{\perp}}$ 
referring only to  solutions of Eq.\ (\ref{MObessel}) 
with $\omega=0$. This method will be very useful in the 
Schwarzschild black--hole case, where the analogue of (\ref{PFRW}) 
cannot be  explicitly obtained.
One considers the form of the solution  of 
Eq.\ (\ref{MObessel}) with arbitrary
frequency
for  large and negative values of $\xi$. This  has a simple 
form:
\begin{equation}
\psi_{\omega k_{\perp}}(\xi) \approx -\frac{1}{\omega}\sin 
[\omega \xi + \alpha(\omega)]  \qquad (\xi < 0, |\xi | \gg 1) , 
\label{Asymptotic} 
\end{equation}
where the normalization constant has been fixed 
to make (\ref{Asymptotic}) compatible with (\ref{aux1}).   
In particular, 
\begin{equation}
\psi_{0 k_{\perp}}(\xi ) \approx - \xi + {\rm const.} 
\qquad (\xi < 0, |\xi | \gg 1) .
\label{aux2}
\end{equation}
By solving Eq.\ (\ref{MObessel})
with $\omega =0$ and fitting the solution obtained to (\ref{aux2}) for 
large and negative  values of $\xi $, we recover (\ref{Zero}).  

In order to calculate the response rate $R^R(k_x,k_y)$ with fixed  
transverse momentum  $(k_x,k_y)$, we 
use the general expression (\ref{Maineq}) identifying  
$U_{0 {\bf s} \lambda} ({\bf x})$ with
$\psi_{0 k_{\perp}}(\xi)e^{ik_x x + ik_y y}/(2\pi)$
[see Eqs.\ (\ref{Udef}), (\ref{Uinpsi}) and (\ref{Zero})].  Thus, we find 
\begin{equation}
R^R(k_x,k_y) dk_x dk_y  = 
\frac{q^2}{4\pi^3 a}[K_0 (k_{\perp}/a)]^2 dk_x dk_y. \label{Resmink}
\end{equation}
The total response rate is obtained by integrating (\ref{Resmink}) 
 over the whole range of transverse momenta as
\begin{equation}
R^R_{\rm tot}=
                  \frac{q^2 a}{4\pi^2}  . \label{Flattotal}
\end{equation}

It is interesting to recall at this point that by a standard
Cartesian--coordinate calculation (see, e.g., Refs.\ \cite{IZ,BD}),  
Eq.\ (\ref{Flattotal}) 
can be shown to be  identical to the emission rate 
of usual Minkowski particles.  
This result is interpreted as follows
\cite{HMS,RW}: {\em The emission of a usual finite--energy 
particle  from a uniformly accelerated source in Minkowski 
vacuum as described by inertial observers corresponds 
to either the {\em emission} or the {\em absorption}
of a zero--energy Rindler particle {\em to} or {\em from} 
the FDU thermal bath as described by uniformly accelerated observers.} 
This is in agreement with Unruh and Wald's inertial interpretation of the
excitation of an accelerated detector \cite{UW}, and with the
discussion of this problem  in terms of classical radiation
\cite{HM}. Although these zero--energy particles are conceptually
well defined, they are not observable by  accelerated observers 
\cite{HMS}. This is compatible with the fact that 
observers coaccelerated  with the source  associate 
no emission of classical radiation
with it \cite{Roh,B}. 

We would like
to call attention  to the fact that we are implicitly 
assuming that the classical source 
is adiabatically switched on and off asymptotically.  Thus, we are not 
concerned with the controversy  as to whether or not there is radiation
from uniformly
accelerated sources which are eternally turned on. 
(See Ref.\ \cite{J} and references therein
for a comprehensive analysis on this issue and Ref.\ \cite{HMS2} 
for a brief discussion of its relation to our problem.)


\section{Static source in Minkowski spacetime}
\label{sec:Minkowski}
 
Before analyzing static sources in the spacetime with a black hole, 
it is useful for later purpose to 
work out the response of a static source in Minkowski  
spacetime using spherical coordinates and assuming a background
thermal bath. 
The line element of Minkowski spacetime in spherical coordinates 
is 
\begin{equation}
ds^2 = dt^2 - dr^2 - 
r^2 (d\theta^2 + \sin^2 \theta d\phi^2)  . \label{MSC}
\end{equation}
In spherical coordinates we write our oscillating source in the form
\begin{equation}
j_{\omega_0}(x) = \frac{\sqrt 2 q \cos \omega_0 t}{r^2\sin\theta}   
\delta (r-r_0) 
\delta (\theta - \theta_0) \delta (\phi - \phi_0) \label{OC0}. 
\end{equation}

Let us write the positive--frequency 
solutions of the massless Klein--Gordon
equation with respect to the Killing 
field $\partial/\partial t$ as
\begin{equation}
u_{\omega l m} = \sqrt{\frac{\omega}{\pi}} {\psi_{\omega l}(r) 
\over r} \times Y_{lm} (\theta, \phi) e^{ -i\omega t} , 
\label{SKG0}
\end{equation}
where $Y_{lm} (\theta, \phi)$ are
spherical harmonics \cite{AS}  
with $l \geq 0$ and $-l \leq m \leq +l$. 
[The form (\ref{SKG0}) 
 will  also be adopted in the following sections.]
Here
$\psi_{\omega l}(r)$ is the 
solution of the ordinary differential equation
\begin{equation}
\left( -\frac{d^2}{dr^2}  + \frac{l(l+1)}{r^2} \right) \psi_{\omega l}(r) 
=  \omega^2 \psi_{\omega l}(r).
\label{RPWE0}
\end{equation}
The solutions of Eq.\ (\ref{RPWE0}) which 
are finite at $r=0$ are 
\begin{equation}
\psi_{\omega l} (r) = C_{\omega l} r j_l (\omega r )  ,
\label{RFM}
\end{equation}
where the $j_l (x)$ are the spherical Bessel  
functions. The normalization constants $C_{\omega l}$ are found 
by substituting  Eq.\ (\ref{SKG0})
in  the Klein--Gordon inner product (\ref{KG}) as 
$$
(\omega +\omega') \int_0^{+\infty} 
dr \psi_{\omega l} (r)^* \psi_{\omega' l} (r)
=
\frac{\pi}{\omega} \delta(\omega-\omega' )
$$
and  using Eq.\ (\ref{RPWE0}) to turn this integral 
into a surface term:
\begin{equation}
\frac{C_{\omega l}^* C_{\omega' l}}{\omega -\omega'} 
\left[
- \omega' r^2\, j_l (\omega r) j_{l+1} (\omega' r)
+ \omega  r^2\, j_l (\omega' r) j_{l+1} (\omega r)
\right]_{r \to +\infty}
=
\frac{\pi}{\omega} \delta (\omega - \omega') ,
\label{KGNSC}
\end{equation}
where use has been made of the following identity 
(see (10.1.22) of Ref.\ \cite{AS}):
$$
\frac{d}{dr} j_l (\omega r) = 
\frac{l}{r} j_l (\omega r) 
-\omega j_{l+1} (\omega r) .
\label{4.6}
$$
In order to evaluate the l.h.s. of  (\ref{KGNSC}) we use
(see (10.1.1) and (9.2.1) of Ref.\ \cite{AS})
\begin{eqnarray}
j_l (x) & = &       \sqrt{\frac{\pi }{2x} } J_{l + 1/2} (x) \nonumber \\
        &\approx & 
        x^{-1} \sin (x - l\pi/2 ) \qquad  (x \gg 1) ,
\end{eqnarray}
and Eq.\ (\ref{Delta}).
Hence, from Eq.\ (\ref{KGNSC}) we obtain $ C_{\omega l} = 1 $ 
(up to a phase), and the
normalized zero--energy mode is 
\begin{equation}
\psi_{0 l} (r) =  r j_l (0) = r\delta_{l0}.
\label{RPKGN}
\end{equation}

In spherical coordinates, a general expression for the response rate with
fixed angular momentum can be obtained by using 
Eq.\ (\ref{SKG0}) in (\ref{Maineq}):
\begin{equation}
R_{l m} =
     \frac{2 q^2 \sqrt{f({\bf x}_0)}}{\beta r_0^2} | \psi_{0 l} (r_0) |^2
     \vert Y_{l m} (\theta_0, \phi_0 )\vert^2 .
\label{above}
\end{equation}
Here $\psi_{0 l}$ is given by (\ref{RPKGN}) and  $f({\bf x}_0) = 1$. 
Thus, we obtain
\begin{equation}
R^{\rm M}_{l m} =
      2 q^2 \beta^{-1}\vert Y_{l m} (\theta_0, \phi_0 )\vert^2\delta_{l0}
= \frac{q^2}{2\pi\beta}\delta_{l0}\,,
\label{AF1A}
\end{equation}
where we have used 
$\vert Y_{00} (\theta, \phi) \vert^2 = (4\pi)^{-1}$.
Since this expression 
vanishes for every $l$ except for $l=0$,
the {\em total} response rate is
\begin{equation}
R^M_{\rm tot}
= \frac{q^2}{2 \pi \beta}\,. 
\label{TRIVIAL}
\end{equation}

One can readily verify that 
the same response rate (\ref{TRIVIAL})
is obtained by repeating the calculation 
in Cartesian coordinates \cite{IZ,BD}.
(In this case the normalized positive--frequency 
modes are the standard ones:
$ u_{\bf k} (x)=  e^{-ik_\mu x^\mu}/ \sqrt{16 \pi^3 \omega}$.)
This should clearly be the case since 
the vacuum is defined through
the same timelike Killing 
field $\partial/\partial t$. 

We note that one obtains Eq.\ (\ref{Flattotal}) by substituting 
$\beta^{-1} = a/2\pi$ in (\ref{TRIVIAL}).
This shows that a uniformly 
accelerated source for a massless scalar field
in Minkowski spacetime responds to the
FDU thermal bath as if it were 
at rest 
in Minkowski spacetime with a background thermal bath
provided that both thermal baths have
the same temperature, as is well known (see, e.g., Ref.\ \cite{BD}).
We will return to this point in Secs.\ \ref{sec:bhS} and \ref{sec:bhS2}. 


\section{Static source outside a toy black hole}
\label{sec:bhtoy}

In this section we will treat a static source in
the spacetime of a black hole with  an artificial gravitational effective
potential simple enough 
to enable us to find the normal modes in terms of well-known 
functions. This will allow us to normalize them for every
frequency $\omega$ and thus we will be able to take the limit 
$\omega \to 0 $
explicitly. We will consider a potential that reproduces the main 
features of the effective potential for the Schwarzschild black hole, 
and compare the results with those
obtained in the next section where we treat the Schwarzschild 
case  using the method outlined in Sec.\ \ref{sec:Rindler}. 
This will provide a useful check for the latter method.

The Schwarzschild line element is 
\begin{equation}
ds^2 = f(r) dt^2 - f(r)^{-1} dr^2 - r^2 (d\theta^2 + \sin^2 \theta d\phi^2) ,
\label{LE}
\end{equation}
where $f(r) = 1-2M/r$. The scalar source $j_{\omega_0}(x)$ is 
given by (\ref{OC0}) with $\sqrt h = f^{-1/2} r^2 \sin \theta$.
The positive--frequency solutions $u_{\omega l m}$
of the massless scalar field equation   
can be written as in Eq.\ (\ref{SKG0})
where 
the $\psi_{\omega l}(r)$ here satisfy the differential equation
\begin{equation}
 \left[ -f(r)
\frac{d}{dr}\left(f(r)
\frac{d\ }{dr} \right) + V_{\rm eff}(r)\right]
\psi_{\omega l}(r)
= \omega^2 \psi_{\omega l}(r),
\label{RPWE}
\end{equation}
with
\begin{equation}
V_{\rm eff}(r) = \left( 1-2M/r\right)
\left[ 2M/r^3 + l(l+1)/r^2\right].
\label{ScV}
\end{equation}
The effective potential $V_{\rm eff}(r)$ vanishes at the 
horizon and goes to zero like $1/r^2$ for large $r$.
It is useful to introduce the
dimensionless Wheeler tortoise coordinate
$x \equiv y +\ln (y-1)$ where 
$y \equiv r/2M$.  Then Eq.\ (\ref{RPWE}) can 
be rewritten as
\begin{equation}
\left[ -{{d^2} \over {d x^2}} + 
 4M^2 V_{\rm eff}[r(x)]\right]\psi_{\omega l} 
= 4M^2\omega^2\psi_{\omega l}. 
\label{KG2}
\end{equation}

There are two 
independent solutions to (\ref{KG2}).  One solution corresponds to the mode
purely incoming from the past horizon $H^-$,  and the other to the mode 
purely incoming from the past null infinity ${\cal J^-}$. These
modes are orthogonal to each other with respect to the Klein--Gordon
inner product (\ref{KG}). 

We will
focus in this and the next sections on the Unruh  vacuum \cite{U} where 
there is a thermal flux of temperature 
$\beta^{-1} = 1/8\pi M$  coming out from $H^-$. 
This is basically
the Hawking radiation \cite{H} which leads to evaporation of
black holes formed by gravitational collapse.   

At this point we will replace the Schwarzschild effective potential
$V_{\rm eff}[r(x)]$ by a simpler potential
\begin{equation}
V_{\rm sim}(x)= \frac{l(l+1)}{4 M^2 x^2} \Theta (x-1)  ,
\label{ToyP}
\end{equation}
where  $\Theta (x)$ is the step function. For $l \neq 0$, this potential
possesses 
the main features of the true potential (\ref{ScV}):
$V_{\rm sim}$ vanishes at the horizon and goes to zero like
$1/r^2$ for large $r$.

For purely incoming waves from the past horizon, $H^-$, 
we write  
\begin{equation}
\psi_{ \omega l} (x) = 
\left\{ 
\begin{array}{l}
A_{ \omega l} (e^{2i M\omega x} + {\cal R}_{ \omega l} 
e^{-2i M\omega x}) 
\ \ \ (x\leq 1) , \label{psiIa} \\
2 i^{l+1} A_{ \omega l} {\cal T}_{ \omega l} 
M\omega x \; h_l^{(1)} ( 2M\omega x) 
\ \ \ (x>1) , \label{psiIb}
\end{array}
\right.
\end{equation}
where $A_{ \omega l}$ is the normalization constant, and 
$|{\cal R}_{ \omega l}|^2$ and $|{\cal T}_{ \omega l}|^2$ are
the reflection and transmission coefficients, respectively. Note that
(see (8.451.3-4) of Ref.\ \cite{GR} and (10.1.1) of Ref.\ \cite{AS})
\begin{eqnarray}
h_l^{(1)} (x) & = & j_l (x) +i n_l (x)  \label{h1h2} \\
              & = & \sqrt{\frac{\pi}{2 x}} H^{(1)}_{l+1/2} (x) \\  
              & = &  (-i)^{l+1} { e^{ix} \over {x}} 
                    [ 1 + {\cal O}(x^{-1}) ]  
                    \qquad  (\vert x \vert \gg  1 ) . \label{AH1}
\end{eqnarray}
 Using the continuity of the modes and their derivatives at
$x=  1$  we find the scattering coefficients:
\begin{equation}
{\cal T}_{ \omega l} =  \frac{e^{2i M\omega }}{i^{l+1} M\omega} [
(1-i/ 2M\omega) h_l^{(1)} ( 2M\omega )-i h_l^{(1)\prime}(2M\omega) ]^{-1} \; ,
\label{betaI}
\end{equation}
and
\begin{equation}
{\cal R}_{ \omega l} =  e^{4i M\omega } 
\frac{[(1+i/ 2M\omega) h_l^{(1)} ( 2M\omega )+i  h_l^{(1)\prime}(2M\omega) ]}
     {[(1-i/ 2M\omega) h_l^{(1)} (2M\omega)-ih_l^{(1)\prime}(2M\omega) ]} \; ,
\label{gammaI}
\end{equation}
where primes indicate  derivatives with respect  
to the argument. From these equations, we 
obtain the usual probability conservation
\begin{equation}
|{\cal T}_{ \omega l}|^2 +|{\cal R}_{ \omega l}|^2 =  1 , 
\label{betagamma1}
\end{equation}
where we have used 
$$
h^{(1)}_l (x) {h^{(1)}_l}^{\prime} (x)^* -  {h^{(1)}_l}(x)^* 
h^{(1)\prime}_l (x) 
=   -2i/x^2 ,
$$
derived from (see (10.1.21-31) of Ref.\ \cite{AS})
$$
j_l'(x) n_l(x) - j_l(x) n_l'(x) =   -1/x^2  .
$$

The normalization constant $A_{\omega l}$ is obtained, as usual,
by substituting
(\ref{SKG0}) in (\ref{KG}) and using Eq.\ (\ref{KG2}) 
with $V_{\rm sim}(x)$ in place of $V_{\rm eff}[r(x)]$
to turn the integral into a surface term
 \begin{equation}
\frac{1}{\omega - \omega'} \left[ {\psi_{\omega' l}} (x)
{d \over {d x}}\psi_{ \omega l}^* (x) - \psi_{ \omega l}^* (x)
{d \over {d x}} {\psi_{ \omega' l}} (x)  \right]^{x \to +\infty}_{x \to
-\infty} 
=  - \frac{2\pi M}{\omega} 
\delta ( \omega -  \omega') ,
\label{again}
\end{equation}
where $\psi_{\omega l} (x)$ is given by (\ref{psiIa}).
Using the large $|x|$ behavior of $\psi_{\omega l}(x)$ obtained by
substituting (\ref{AH1}) in (\ref{psiIa}), 
we obtain (up to  a phase) from (\ref{again})
\begin{equation}  
 A_{ \omega l}  =  (2\omega)^{-1} .
\label{A}
\end{equation}
Substituting the scattering coefficients  
(\ref{betaI})--(\ref{gammaI}) and the normalization constant (\ref{A}) in 
Eq.\ (\ref{psiIa}), we obtain
\begin{equation}
\psi_{ 0  l} (x)  =
\left\{
\begin{array}{l}
2 M (l^{-1} +1 - x) \qquad (x \leq 1) \label{psiISka}  ,\\
2 M l^{-1} x^{-l} \qquad (x > 1) ,
\end{array}
\right.
\end{equation}
where we have used  
\begin{equation}
 j_l (x)  = c_l x^l +{\cal O} (x^{l+2}) ,
\qquad 
n_l(x) = - d_l x^{-l-1} - e_l x^{-l+1} +{\cal O} (x^{-l+3})
\label{jn}
\end{equation}
with $c_l=  1/ (2l+1)!!$,  $d_l=   (2l-1)!!$, and $e_l =(2l-3)!!/2$.
Note here that $\psi_{0l}(x) = -2Mx + {\rm const.}$ for $x < 0$.

In order to calculate the response rate, we use (\ref{above}) with $\beta^{-1}
=1/8\pi M$.  
In the case the source is at $x_0 \leq 1$ ( $ r_0 /2M \leq 1.567 $)
we obtain 
\begin{equation}
R^{\rm T}_{l m}(x_0) =
           {q^2 M \over {\pi}} 
           {f(r_0)^{1/2} \over{r_0^2}} 
            \left\{
           l^{-1}  - y(r_0) f(r_0)  - 
           \ln \left[ y(r_0) f(r_0) \right]
            \right\}^{2} | Y_{l m} (\theta_0, \phi_0)|^2 ,\ \ x_0 \leq 1,
\label{RF1}
\end{equation}
where $y(r) = r/2M $, 
while in the case the source is at
$x_0 > 1$ ( $ r_0/2M > 1.567 $) we obtain
\begin{equation}
R^{\rm T}_{l m}(x_0)=
           {q^2 M \over {\pi }} 
           {f(r_0)^{1/2} \over{r_0^2 \, l^2}} 
           \left\{
                 y(r_0)  + \ln \left[ y(r_0) f(r_0)  \right]
           \right\}^{-2l} |Y_{l m} (\theta_0, \phi_0)|^2 ,\ \ x_0 >1,
\label{RF2}
\end{equation}
where $l \neq 0$. 
Note here that for $x_0 \gg 1$ we have 
\begin{equation}
R^T_{lm} (x_0) \approx (q^2/4\pi M) l^{-2} x_0^{-2l-2}|Y_{lm}|^2.
\label{MODEL}
\end{equation}


\section{Static source outside the Schwarzschild black hole with the
Unruh vacuum}
\label{sec:bhS}

The fact that must be considered first in solving the full Schwarzschild
case is that very little is known about the 
solutions of the wave equation (\ref{KG2}) of nonzero frequency $\omega$
with potential (\ref{ScV}). (See Ref.\ \cite{JC} for some known properties of
these solutions.) 
Thus  we use the method outlined in Sec.\ \ref{sec:Rindler} [see the 
paragraph below
Eq.\ (\ref{Zero})] 
in order to normalize the zero--energy
modes which are the  only relevant ones here [see Eq.\ (\ref{above})].
We will consider the Unruh vacuum as in the last section. Thus,
we  need to consider only the modes incoming from $H^-$. Close to and far
away from the horizon we can  write 
\begin{equation}
\psi_{ \omega l} (x) \approx 
\left\{ 
\begin{array}{l}
A_{ \omega l} (e^{2i M\omega x} + {\cal R}_{ \omega l} e^{-2i M\omega x}) 
\ \ \  (x<0,\ |x| \gg 1) ,\\
2i^{l+1} 
A_{ \omega l} {\cal T}_{ \omega l} \; 
M\omega x \; h_l^{(1)} ( 2M\omega x) 
\ \ \ (x \gg 1) . 
\end{array}
\right.
\end{equation}
Zero--frequency modes coming from $H^-$ 
are totally reflected back by the potential toward the horizon
[see Eq.\ (\ref{psiISka}) with $x \gg 1$]. 
This implies that
for $M\omega \ll 1$  the behavior of $\psi_{\omega l}(x)$ close
to the horizon determines 
the inner--product 
in (\ref{KG}).  Taking this fact into account  and disregarding 
the black--hole potential close to the horizon, we find the normalized
solution of Eq.\ (\ref{KG2}) in this region:
\begin{equation}
\psi_{\omega l} (x) \approx -\omega^{-1} 
\sin [2M\omega x + \alpha(\omega)] \ \ \ \ (x < 0,\ |x| \gg 1)
\end{equation}
up to a phase.
In the limit $\omega \to 0$, we find
\begin{equation}
\psi_{0l} (x) \approx - 2M x + {\rm const.} \ \ \ \ (x < 0,\ |x| \gg 1).
\label{Norma}
\end{equation}
This agrees with the behavior 
of $\psi_{0l}(x)$ for large and negative $x$ found in the last section  
by normalizing the modes of nonzero $\omega$ and then taking the limit
$\omega \to 0$ [see Eq.\ (\ref{psiISka})].

We note that  for $\omega = 0$
Eq.\ (\ref{KG2}) can be reduced to the Legendre equation.
The general solution is \cite{C}
\begin{equation}
\psi_{0l}(y) = C_1 yQ_l(2y-1) + C_2 yP_l(2y-1),
\label{gene}
\end{equation}
where $y=r/2M$, and
$P_l(z)$ and $Q_l(z)$ are the Legendre functions 
of the first and second kinds
with the branch cut $(-\infty, 1]$ for $Q_l(z)$. 
By recalling that for $\omega \to 0$ the solution we seek  must 
be totally reflected back to the horizon, and that $P_l(z)\sim z^l$ and 
$Q_l(z)\sim z^{-l-1}$ for large $z$, we  conclude that we must 
let $C_2 = 0$. We find the normalization constant $C_1$ by comparing  
(\ref{gene}) 
close to the horizon with (\ref{Norma}). For this purpose, note that 
(see (8.834.2) and (8.831.3) of  Ref.\ \cite{GR})
$$
Q_l(z) = \frac{P_l(z)}{2} \ln\frac{z+1}{z-1} - W_{l-1} (z)
$$
where
$$
W_{l-1} (z) = \sum_{k=1}^l k^{-1} P_{k-1}(z) P_{l-k} (z) .
$$
Thus from Eq.\ (\ref{gene}), we have 
\begin{equation}
\psi_{0l}(x) \approx  
C_1  \left(  \frac{1}{2} - \frac{x}{2} -\sum_{m=1}^l \frac{1}{m} \right) 
\ \ \ \ (x < 0,\ |x| \gg 1) .
\label{Ver}
\end{equation}
By comparing (\ref{Ver})  with (\ref{Norma}) we obtain $C_1 = 4M$.
Thus
\begin{equation}
\psi_{0l} (x) = 4MyQ_l(2y-1) .
\label{ola}
\end{equation}

The response rate to quanta of given angular momentum is 
readily obtained by substituting
(\ref{ola}) in (\ref{above}):
$$
R_{lm}^{S-U} = \frac{q^2}{\pi M}f(r_0)^{1/2}
|Q_l(z_0)|^2 |Y_{lm}(\theta_0,\varphi_0)|^2,
$$
where $z_0 \equiv r_0/M-1$.
Note that for $x_0 \gg 1$ we have 
$$
R^{S-U}_{lm}(x_0)\approx (q^2/4\pi M)[(l!)^2/(2l+1)!]^2(x_0)^{-2l-2}|Y_{lm}|^2.
$$  
Comparison of this equation and Eq.\ (\ref{MODEL}) shows that 
the rate obtained
with the toy black hole does model 
the exact response rate for moderate $l$ provided that the source is
set at large $x_0$. 
In order to obtain the total response rate $R_{\rm tot}^{S-U}$, 
we sum over $l$ and $m$. For this purpose we use
$$
\sum_{m=-l}^{l}|Y_{lm}(\theta,\varphi)|^2 = \frac{2l+1}{4\pi}
$$ and 
$$
\sum_{l=0}^{\infty}(2l+1)[Q_l(z)]^2 = \frac{1}{z^2 -1}\, .
$$
The latter expression can be obtained by squaring the formula
$
\sum_{l=0}^{+\infty}P_l(t)Q_l(z) = (z-t)^{-1} ,
$
and integrating from $-1$  to $1$ with respect to $t$.
In this way we obtain 
\begin{equation}
R^{S-U}_{\rm tot}
= \frac{q^2 a(r_0)}{4\pi^2}, \label{RS-U}
\end{equation}
where $a(r_0) = Mf(r_0)^{-1/2}/r_0^2$ is the proper acceleration of the 
static source. 
Note that Eq.\ (\ref{RS-U}) is identical to (\ref{Flattotal}) as a
function of proper acceleration. This is our main result:
{\em The emission and absorption of zero--energy particles by 
a static source outside a Schwarzschild black hole with the initial 
quantum state being the Unruh vacuum
is exactly
the same as if the source were static in the Rindler wedge with 
the initial quantum state being the Minkowski vacuum.}
Note that close to the horizon Eq.\ (\ref{RS-U}) can be written 
as a function
of the proper temperature (see Ref. \cite{T}), 
$\beta^{-1} = f^{-1/2}/ 8\pi M$, as
\begin{equation}
R^{S-U}_{\rm tot}
\approx \frac{q^2}{2 \pi \beta} \qquad (x <0,\ |x| \gg 1), 
\end{equation}
which is the same as (\ref{TRIVIAL}). 


\section{Static source outside a Schwarzschild black hole with 
         the Hartle--Hawking vacuum}
\label{sec:bhS2}

In this section 
we will calculate
the response rate of the static source  when the initial state is taken
to be the Hartle--Hawking vacuum \cite{HH}.
(This will show that the Unruh vacuum state is essential for the above
mentioned equality.)
In this state thermal fluxes come in from ${\cal J}^-$ as well as
from $H^-$.  The contribution of the flux from $H^-$ to the response rate
has already been calculated in the previous section. 
Thus, we consider here the modes incoming
from ${\cal J}^-$.
Close to and far away from the horizon these modes can be written  as
\begin{equation}
\psi_{ \omega l} (x) \approx 
\left\{
\begin{array}{l}
A_{ \omega l} {\cal T}_{\omega l} e^{-2iM \omega x}
\qquad (x<0,\ |x| \gg 1) ,\label{psiIIa} \\
A_{ \omega l} [ 2(-i)^{l+1} M \omega x \; {h_l^{(1)}} (2M \omega x)^* 
 + 2i^{l+1}{\cal R}_{ \omega l} \; M \omega x \; h_l^{(1)} (2M \omega x) ] 
\qquad (x \gg 1) ,\label{psiIIb} 
\end{array}
\right.
\end{equation}
where the normalization 
constant can be determined by the procedure used in Sec.\ \ref{sec:bhtoy}
with the same result
$A_{\omega l} = (2\omega)^{-1}.$

Recall that Eq.\ (\ref{gene}) gives the general expression for the
zero--frequency solution $\psi_{0 l} (x)$ of (\ref{KG2}). 
Because zero--frequency modes must be totally reflected 
by the  black--hole potential toward ${\cal J}^+$, we  conclude that 
in this case $C_1=0$. 
(Note that $P_l (1) = 1$ while $Q_l(z) \approx - \log \sqrt{z-1}$ for 
$z \approx 1$ and that $P_l(z)$ and $Q_l(z)$ behave like $z^l$ and
$z^{-l-1}$, respectively, for $z \gg 1$.)
Thus
\begin{equation} 
\psi_{0l}[x(y)] = C_2 yP_l(2y-1) ,
\label{gene'}
\end{equation}
where $x(y) = y + \ln(y-1)$.
In order to determine the normalization constant $C_2$ we
 first note that for large $x$ 
Eq.\ (\ref{gene'}) can be written  as 
(see (8.820), (8.837.2) and (8.339.2) in Ref.\ \cite{GR})
\begin{eqnarray}
\psi_{0 l} (x) 
&\approx & C_2 x F(-l,l+1;1;1-x) \qquad (x \gg 1) \nonumber \\
&\approx & C_2 \frac{(2l+1)!}{l!^2} x^{l+1} \qquad (x \gg 1) .
\label{gene'as}
\end{eqnarray}
Now we find from (\ref{psiIIb})  the following expression 
for $1\ll x \ll 1/M\omega$ [see Eqs.\ (\ref{h1h2}) and (\ref{jn})]:
\begin{equation}
\psi_{ \omega l} (x) 
\approx \frac{2^{2l+1} l! M^{l+1} \omega^l }{(2l+1)!}x^{l+1}  \qquad 
(1 \ll x \ll 1/M\omega),
\label{psiIIb'}
\end{equation}
where we have set ${\cal R}_{\omega l} \approx (-1)^{l+1}$ for $\omega \ll 1$
so that $\psi_{\omega l}(x)$ behaves like $x^{l+1}$ in the specified
range of $x$.
By comparing (\ref{gene'as}) and (\ref{psiIIb'}),  we obtain
\begin{eqnarray}
C_2 & = & \left. 
       \frac{2^{2l+1} (l!)^3 M^{l+1} \omega^l}{(2l +1)!^2} 
      \right|_{\omega \to 0} 
      \nonumber \\
    & = &  2 M \delta_{l 0}\,.
\label{C1norm}
\end{eqnarray}
Thus the only non--vanishing zero--energy mode $\psi_{0 l} (x)$ has 
vanishing angular momentum.

 Using Eq.\ (\ref{above}) with (\ref{gene'}) and (\ref{C1norm}),
we write the total contribution to the response rate due to
the modes incoming from ${\cal J}^-$ in terms of the 
proper acceleration as
\begin{equation} 
R_{{\cal J}^-} = \frac{q^2}{16 \pi^2 r_0^2 a(r_0)} .
\label{contribution}
\end{equation}
Therefore the total response rate of our scalar source in the 
Hartle--Hawking vacuum  is given by the sum of (\ref{contribution})
and (\ref{RS-U}):
\begin{equation}
R^{S-HH}_{\rm tot}
= \frac{q^2 a}{4\pi^2}  + \frac{q^2}{ 16 \pi^2 r^2_0 a }  .
\label{RS-HH}
\end{equation} 
Clearly Eq.\ (\ref{RS-HH}) differs from Eq.\ (\ref{Flattotal})
by the second term on the r.h.s. 
Thus, the equality found for the Unruh vacuum
does not hold for 
the Hartle--Hawking vacuum.
(However, if we considered the massless limit of a massive field, the second
term in Eq.\ (\ref{RS-HH}) would be absent. Hence, the equality would hold.) 
Note that,
for large $x$, Eq.\ (\ref{RS-HH}) can be written as a function
of the proper temperature as
\begin{equation}
R^{S-HH}_{\rm tot}
\approx \frac{q^2}{2 \pi \beta} \qquad (x \gg 1) , 
\end{equation}
which agrees with (\ref{TRIVIAL}). 
This is consistent with the fact that 
far away from the black hole the Hartle--Hawking
vacuum is identical with a thermal bath in Minkowski spacetime.


\section{Consistency with the literature}
\label{sec:literature}

Before showing that our results are in agreement with  those of 
Refs.\ \cite{C,CSD} 
by Candelas and Sciama et al. (CSD),
we recall (i) that our response rate 
is the {\it sum} of absorption and
emission rates of zero--energy modes, and (ii) 
that the absorption and
emission rates of zero--energy modes are equal. Thus 
our result would be twice the absorption rate
obtained by CSD in the zero--energy limit  {\it if 
we used the same source}. However, this is not the case.
The classical source equivalent to the detector 
in Refs.\ \cite{C,CSD} is proportional to 
$\exp (i\omega_0 t)$.  This is replaced in our case by
$\sqrt 2 \cos (\omega_0 t)= \exp (i\omega_0 t)/\sqrt2 
                          + \exp (-i\omega_0 t)/\sqrt2$.
Since the second term on the r.h.s. does not contribute to the absorption
rate in the computation, our source   
is effectively $1/\sqrt{2}$ of the one
in Refs.\ \cite{C,CSD} as far as absorption is concerned. 
Eventually, when we square 
the amplitude to obtain 
the probability, we end up with an {\em absorption} 
rate which is $1/2$ of 
the one obtained by CSD in the $\omega_0 \to 0$ limit. 
Hence, our results will be 
compatible with CSD if our total 
response ({\em absorption} + {\em emission}) rate 
is equal to the absorption rate 
of Refs.\ \cite{C,CSD} in the $\omega_0 \to 0$ limit.
 
To show that this is indeed the case it is convenient to interpret our
results (with $q^2 = 1$) in terms of the two--point function as follows.
Let us consider a 
static world line and let $x_\tau$ be  
the spacetime point which corresponds to proper time
$\tau$ measured along the world line of the scalar source.  
Then the rate (per proper time) we have computed 
in Schwarzschild spacetime with the 
field in the {\em Unruh vacuum}
initial state, $|0\rangle_U$, is
$$
R^{S-U}_{\rm tot} = 
\int d\tau\,\, _U\langle 0|\phi(x_\tau)\phi(x_0)|0\rangle_U\,,
$$
where $_U\langle 0|\phi(x)\phi(y)|0\rangle_U$ is the two-point function
of the massless scalar field $\phi$.
The behavior of this quantity near and far away from the horizon can be
found using the results in CSD as follows.
 
The quantity considered by CSD is
$$
\Pi(\omega|r) = \int_{-\infty}^{+\infty}
dt\, {\rm exp}(-i\omega t)\,\,_U\langle 0|
\phi(x(t))\phi(x(0))|0 \rangle_U ,
$$
where $t$ is the {\em coordinate} time along a static world line.
Since $d\tau = (1 - 2M/r)^{1/2} dt$, our response rate 
$R^{S-U}_{\rm tot}$ 
and $\Pi(0|r)$ should be related by
$$
R^{S-U}_{\rm tot} = (1-2M/r)^{1/2}\Pi(0|r).
$$
Now, according to CSD, for $r \approx 2M$ one has
$$
(1-2M/r)^{1/2}\Pi(\omega|r) \approx
\frac{\omega}{2\pi(1-2M/r)^{1/2}[{\rm exp}(2\pi \omega /\kappa) -1]},
$$
where $\kappa = 1/4M$.  Taking the limit $\omega \to 0$ for 
$r \approx 2M$, one finds
\begin{eqnarray}
(1-2M/r)^{1/2}\Pi(0|r) &\approx & \frac{1}{16\pi^2 M (1-2M/r)^{1/2}}
\nonumber \\
&\approx & \frac{1}{4\pi^2}\frac{M}{r^2}(1-2M/r)^{-1/2} .
\nonumber 
\end{eqnarray}
This is in agreement with our formula (\ref{RS-U}) with
$q^2 =1$.

The calculation for  large values of $r$ is a little 
more involved.
The formula given by CSD for this case is
$$
\Pi(\omega|r)
\approx \frac{1}{r^2}
\frac{\Sigma_{l=0}^{+\infty} (2l+1)|B_l(\omega)|^2}{8\pi\omega
[{\rm exp}(2\pi\omega/\kappa) -1]}-\frac{\omega}{2\pi}\theta(-\omega).
$$
(The factor $(1-2M/r)^{1/2}$ is irrelevant now.)   
Taking the $\omega \to 0$
limit, one has
\begin{equation}
\Pi(0|r)
\approx \frac{1}{64\pi^2 M r^2}
\sum_{l=0}^{+\infty}(2l+1)\lim_{\omega \to 0} 
\frac{|B_l(\omega)|^2}{\omega^2},
\label{help2}
\end{equation}
where the $|B_l(\omega)|^2$ are the transmission coefficients 
for the modes incoming from $H^-$.
In our notation where the positive--frequency modes are given by 
Eq.\ (\ref{SKG0}), the $B_l(\omega )$ are defined by
\begin{equation}
\psi_{\omega l} (r) \approx 
\left\{ 
\begin{array}{l}
C_{\omega l} (e^{2iM\omega x} + A_l(\omega ) e^{-2iM\omega x} ) 
\ \ \ (r \approx 2M) , \\
2i^{l+1}C_{\omega l} B_l(\omega ) \; M\omega x \; h_l^{(1)} (2M \omega x) 
\ \ \ (r \gg 2M) ,
\label{help}
\end{array} \right.
\end{equation}
where $C_{\omega l} =(2\omega)^{-1}$ [see Eq.\ (\ref{A})]. Note that for
$1 \ll x \ll 1/M\omega$ we have from (\ref{help})
\begin{eqnarray}
\psi_{\omega l} (r) 
& \approx & B_l(\omega ) i^{l+1} 
Mx [j_l (2M\omega x) +i n_l (2M\omega x)]
\qquad (x \gg 1) \nonumber \\
& \approx & \frac{B_l(\omega )}{\omega} 
\frac{i^{l+1} (2l)!\, \omega^{-l}}{2^{2l+1} l! M^l} x^{-l} 
\qquad (1 \ll x \ll 1/M\omega) .
\label{psilarge1}
\end{eqnarray}

In order to determine $B_l(\omega )$ for small $\omega$, 
we recall that the normalized 
zero--frequency solution of Eq.\ (\ref{KG2}) can be written as 
[see Eq.\ (\ref{ola})] 
\begin{eqnarray}
\psi_{0l} (r ) &=&  4MyQ_l(2y-1) \nonumber \\
              &=&   
   4My \frac{\Gamma(l+1) \Gamma (1/2)}{2^{l+1} \Gamma(l+3/2)} 
   F \left( \frac{l+2}{2}, \frac{l+1}{2}; \frac{2l+3}{2};(2y-1)^{-2} \right)
      \frac{1}{(2y-1)^{l+1}}  \nonumber \\       
              &\approx & \frac{2M(l!)^2}{(2l+1)!} x^{-l} \qquad (x \gg 1) ,
\label{psilarge2}
\end{eqnarray}
 where $y=r/2M$.
 Thus, comparing (\ref{psilarge1}) and (\ref{psilarge2}) we have
\begin{eqnarray}
\left. \frac{B_l(\omega)}{\omega} \right|_{\omega \to 0}
&=& 
\left. 
\frac{2^{2l+2} (l!)^3 M^{l+1} \omega^l}{(2l)! (2l+1)!}
\right|_{\omega \to 0} \nonumber \\
&=& 4M 
\delta_{l0}\,. 
\label{help3}
\end{eqnarray}
Finally,  substituting (\ref{help3}) in (\ref{help2}),
we  recover the large $r$ limit of our formula (\ref{RS-U}) with $q^2=1$.
Thus, our results are in agreement with CSD for $r\approx 2M$
and for large $r$.


\section{Discussions}
\label{sec:discussions}

We have shown that there is an equality between 
the response of a static source in Schwarzschild spacetime
(with the Unruh vacuum) and that of a uniformly accelerated source 
in Minkowski spacetime (with the Minkowski vacuum)
provided that the proper acceleration is the same. This result was quite
unexpected since all classical formulations of the equivalence 
principle are valid only locally while quantum states are defined  
globally.  What could have been naturally expected is an
equivalence in the response of the type mentioned 
above  only close to and far away from the horizon 
(see, e.g., Refs.\ \cite{FH,KW}) rather than everywhere.
We have also verified that
close and far away  from the horizon  the  source  responds 
to Hawking radiation as if it were at rest
in a thermal bath in Minkowski spacetime 
characterized by the same proper  temperature
in the Unruh and Hartle--Hawking vacua, respectively.
Clearly, Hawking radiation was crucial in obtaining non--vanishing rates:
Had we chosen the Boulware vacuum \cite{Boul}, we would have obtained 
vanishing response rates. 
It was also shown that the equality derived for the Unruh vacuum does not
hold for the Hartle--Hawking vacuum. (The corresponding result for the 
electromagnetic field, which {\em does not} exhibit any equality, will be
presented elsewhere \cite{CHM}.) 
The procedure used in Schwarzschild
spacetime to normalize  the massless Klein--Gordon scalar field in 
the zero--frequency limit was checked by comparing 
it with the one performed for a toy black hole
where the normal modes can be written  explicitly  for every frequency.
Finally, our results were compared with the literature 
and shown to be in agreement with it.


\acknowledgments

We thank Bob Wald and Bernard Kay for useful discussions.
We also thank Chris Fewster 
for helpful comments 
on the zero--energy limit of one--dimensional scattering theory.
The work of AH was supported in part by 
Schweizerischer Nationalfonds and the Tomalla Foundation.
GM would like to acknowledge partial support from
the Conselho Nacional de Desenvolvimento Cient\'\i fico e
Tecnol\'ogico.
DS would like to acknowledge partial support from 
DGAPA--UNAM Project No. IN 105496.


\end{document}